\documentclass[12pt]{article}

\usepackage{scicite}

\usepackage{times}
\usepackage{graphicx}
\usepackage{amssymb}
\usepackage{amsthm}
\usepackage{amsmath}
\usepackage{soul,color}
\usepackage{lineno}
\usepackage{float}
\usepackage{hyperref}
\usepackage{tabularx}
\restylefloat{table}
\newcolumntype{Y}{>{\centering\arraybackslash}X}

\title{Deep learning enabled laser speckle wavemeter with a high dynamic range}

\author
{Roopam K. Gupta,$^{1,2\ast}$ Graham D. Bruce,$^{1}$ Simon J. Powis,$^{2}$ Kishan Dholakia$^{1,3}$\\
\\
\normalsize{$^{1}$SUPA, School of Physics and Astronomy, University of St. Andrews, KY16 9SS, UK}\\
\normalsize{$^{2}$School of Medicine and Biomedical Sciences Research Complex,}\\
\normalsize{University of St. Andrews, KY16 9TF, UK}\\
\normalsize{$^{3}$Department of Physics, College of Science, Yonsei University, Seoul 03722, South Korea.}
\\
\normalsize{$^\ast$Corresponding Author: rg211@st-andrews.ac.uk}
}

\date{}

\begin{document} 


\baselineskip24pt


\maketitle


\begin{abstract}
 The speckle pattern produced when a laser is scattered by a disordered medium has recently been shown to give a surprisingly accurate or broadband measurement of wavelength. Here it is shown that deep learning is an ideal approach to analyse wavelength variations using a speckle wavemeter due to its ability to identify trends and overcome low signal to noise ratio in complex datasets. This combination enables wavelength measurement at high precision over a broad operating range in a single step, with a remarkable capability to reject instrumental and environmental noise, which has not been possible with previous approaches. It is demonstrated that the noise rejection capabilities of deep learning provide attometre-scale wavelength precision over an operating range from 488~nm to 976~nm. This dynamic range is six orders of magnitude beyond the state of the art.
\end{abstract}


\section{Introduction}
A key property of monochromatic optical waves is their wavelength. An accurate measurement of wavelength can enable many studies in fluorescence spectroscopy, atomic physics, and high precision metrology~\cite{PhysRevA.92.052501,erf2012speckle}. A standard wavemeter has a limitation of a one dimensional dispersion. Using a dispersive element such as a diffraction grating can provide high bandwidth but resolution scales linearly with system size, whereas stabilised Fabry-Perot cavities can obtain high precision over a narrow operating range~\cite{Chakrabarti2015}. Recently, it has been recognised that speckle, which is the granular interference pattern produced when light propagates through a disordered medium, can overcome these limitations by multiplexing spatial-to-spectral mapping in a compact system. Tracking changes in this speckle pattern allows wavelength measurement with high resolution \textit{or} broad operating range~\cite{cao2017perspective}. For wavelength measurement, speckle may be generated by passing light through either a multi-mode fibre~\cite{Redding2012,Redding2013,Redding2014,Wan2015,Bruce:19,Bruce2020} or into an integrating sphere~\cite{Metzger2017,ODONNELL2020124906}.

Capturing the speckle is insufficient: the critical step relies on the interpretation and understanding of the variation in speckle pattern as a function of wavelength. Without the detailed knowledge of the details of laser beam and medium, it is not generally possible to write a mathematical expression for the dependence of the speckle pattern on the wavelength. Instead, the extraction of wavelength must be accomplished by a data--driven approach in which a training phase uses a set of speckle images obtained at known wavelengths to identify the dependence of speckle on wavelength. Most of the previous applications use a method of  calculating the transmission  matrix of the given disordered medium~\cite{PhysRevLett.104.100601}, and this gives the capability to measure wavelength over a range restricted solely by the sensitivity range of the camera used to image the speckle. However, the resolution of this method is limited by a high degree of correlation between the speckle patterns produced at closely-separated wavelengths~\cite{Mazilu2014}, typically on the picometre-scale. Multivariate analysis, in particular principal component analysis (PCA), adds a new dimension to this methodology and has allowed attometre-resolved wavelength measurements~\cite{Bruce:19}. However, PCA allows for only a limited range of operation that has been demonstrated to be, at most, 5 orders of magnitude higher than the resolution~\cite{Metzger2017}. Whilst these results are impressive, to truly convert speckle into a very precise measurement, a powerful single-step algorithm would be desirable. The speckle patterns generated after transmitting light through a disordered medium are unique for each wavelength, with the presence of inherent noise due to environmental or instrumental variations. As a consequence, speckle is likely to be an ideal candidate for the training of a deep learning based classification method~\cite{6472238}.

Deep learning is a powerful technique which has provided major advances in many areas of sciences, from evolutionary biology to quantum physics~\cite{10.1126/sciadv.aaw4967,PhysRevLett.122.040504}. Particularly,
deep learning based artificial neural networks (ANNs) automatically learn to identify and extract the relevant features present in an input dataset~\cite{Gupta:19}. Moreover, the methodology for the application of ANNs makes them universal functional approximators~\cite{Hornik1989} which are widely applied across physical sciences~\cite{Baldi2014,Rem2019,wang2019deep}. Deep learning based convolutional neural networks (CNNs) have already found application in speckle analysis for imaging applications
~\cite{Horisaki2018,Li2018}. Of particular relevance here, they have also been implemented to discriminate between different speckle-creating scatterers~\cite{Valent2018}. Additionally, harnessing the spectral characteristics of speckle, CNNs have found an application to achieve real-time recovery of hyperspectral information with a wavelength resolution of 5~nm~\cite{Kurum:19}.

In this study, we present a method based on deep learning and t-distributed stochastic neighbor embedding (t-SNE)~\cite{Maaten2008} to classify and segment the speckle images corresponding to a given laser wavelength. An interesting aspect presented in this study is the automatic rejection of instrumental or environmental noise by the CNN, which enables a classification of speckle patterns with a wavelength precision of two attometres, representing a nine orders of magnitude improvement compared to previous studies with deep learning~\cite{Kurum:19}. This, coupled with the capability of the pre-trained CNN to segment the speckle images covering the entire visible spectrum, leads to a dynamic range improvement by six orders of magnitude. Going beyond the capability to identify the speckle-creating scatterer~\cite{Valent2018}, we additionally show that the trained CNN, in combination with t-SNE, can recognize the wavelength variations of speckle regardless of which scattering medium is used.

\section{Methods}
\label{methods}
\begin{figure}[h]
  \centering
  \includegraphics[width=1\linewidth]{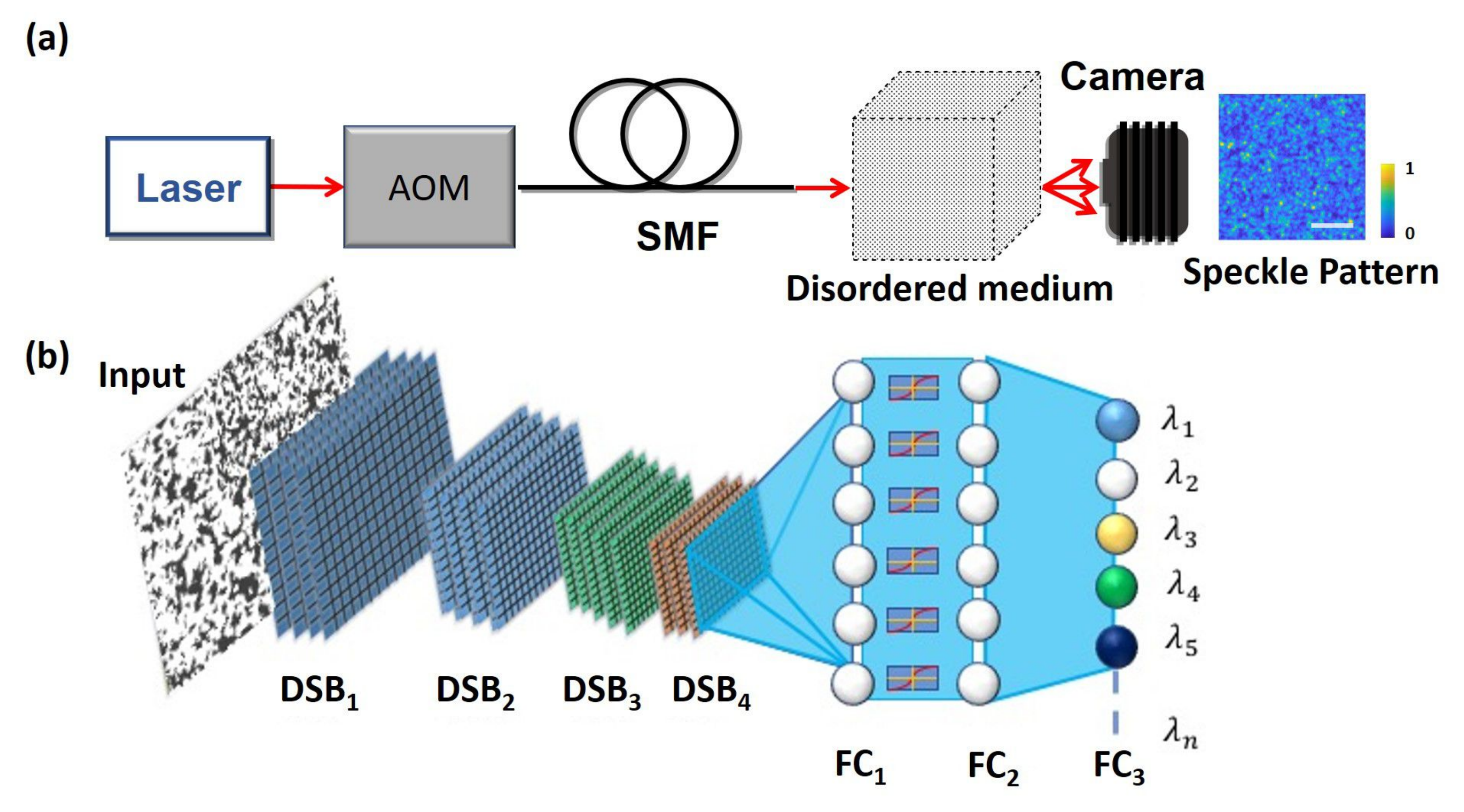}
  \caption{Speckle wavemeter assembly and CNN geometry. (a) The experimental assembly for a speckle wavemeter. The laser wavelength is set using an acousto-optic modulator (AOM) and injected into the disordered medium via a single mode fibre (SMF). The output speckle pattern is captured by the camera. (b) The convolutional neural network (CNN) used to classify the speckle images with respect to the incident laser wavelengths. The CNN consists of an input layer, multiple down-sampling blocks (DSB$_i$) and three fully connected layers (FC$_i$). Here $\lambda_i$ denotes the output wavelength class. The white scale bar on the representative speckle pattern represents 224~$\mu$m, while the intensity is normalized as shown in the adjacent color bar.}
  \label{fig:Fig-1}
\end{figure}

The principle of our approach to measuring wavelength is outlined in Fig.~\ref{fig:Fig-1}. We record on a camera the speckle patterns produced by scattering laser light from a disordered medium (Fig. \ref{fig:Fig-1} (a)). Unless stated otherwise, we use a tunable diode laser which is wavelength-locked to a rubidium reference ($\sim$ 780~nm) as our source of laser light, an acousto-optic modulator to apply controlled wavelength variations, and an integrating sphere to scatter light. 

To extract the wavelength dependence of the accumulated speckle images we implemented a supervised deep learning based convolutional neural network (depicted in Fig. \ref{fig:Fig-1} (b)).

\subsection{Data Acquisition}
\label{DataAcquisition}
The speckle images corresponding to the incident laser wavelength were generated by using a 1.5 inch diameter, spectralon integrating sphere. The laser light from an external cavity diode laser (Topica DL-100 / LD-0785-P220) was stabilized to the $^{87}$Rb D2 line ($F=2 \rightarrow F \textquoteright =2\times3$ crossover at $\sim$ 780~nm) using saturated absorption spectroscopy and top-of-fringe locking. The light from the laser was passed through an acousto-optic modulator (AOM) (Crystal Technologies 3110-120) in a cat-eye double pass configuration to control the wavelength. Speckle is sensitive to many other laser parameters, including the polarisation \cite{Facchin20} and the transverse mode profile of the beam \cite{Mourka13,reddy2014higher,hu2020does}. To ensure the variations in the speckle arise only from wavelength changes, the light was linearly polarised using a polarising beam splitting cube. To remove any variations in the spatial beam profile, the light was coupled into an angle cleaved single-mode fibre (SMF) (Thorlabs P5-780M-FC-10). This was connected to the integrating sphere input-port via an FC/PC connector without collimation optics to produce a diverging fundamental Gaussian mode within the integrating sphere. The SMF delivered 900~$\mu$W into the integrating sphere. The highly Lambertian diffusive coating and multiple reflections create large optical path differences allowing a high resolution for the system.~As the generated speckle pattern also depends on the choice of observation plane, the light then propagated for a fixed distance of 20~cm before impinging the CMOS camera (Mikrotron EoSens 4CXP). This distance was chosen to achieve fully-developed speckle patterns with a mean grain size of $\sim$3~pixels (Fig. \ref{fig:Fig-1} (a)) to prevent sub-Nyquist sampling and associated aliasing effects.

To test the wavemeter over a broader range of the optical spectrum, we use additional lasers at wavelengths of 488~nm (M-Squared frequency-doubled SolsTis Ti:Sapphire), 532~nm (Oxxius single-longitudinal mode diode-pumped solid state laser), 671~nm (Thorlabs HL6756MG Diode Laser) and 976~nm (M-Squared SolsTis Ti:Sapphire). In order to test the generalization capabilities of the CNN, we also performed experiments where the integrating sphere was replaced with a ground glass diffuser (Thorlabs ED1-S20).

During the data accumulation, we recorded a continuous train of 10,000 $128 \times 128$ pixel speckle images for each wavelength at a frame rate of 1 kHz with an exposure of 998~$\mu$s which took a total of 10 seconds. The time difference between the data accumulation of the different wavelength classes was typically 0.5 seconds.

\subsection{Deep learning model architecture, training and calibration}
\label{DLTraining}
To extract the wavelength dependence of the accumulated speckle images we implemented a supervised deep leaning based convolutional neural network (CNN) (depicted in Fig. \ref{fig:Fig-1}(b)). The implemented CNN architecture consists of 4 down-sampling blocks (DSB). Each block consists of 3 convolution layers with 30 filters. Each convolution layer is followed by a batch normalization layer and a rectifier linear unit  (ReLU) activation function layer. To systematically reduce the dimensionality of the input image, each DSB is connected with a max pooling layer with filter size of 2 px $\times$ 2 px. The filter sizes of convolution layers vary as 5 px $\times$ 5 px, 4 px $\times$ 4 px and 3 px $\times$ 3px respectively with a stride and padding of 1 px $\times$ 1 px. The DSBs are followed by two fully connected (FC) layers with leaky ReLU~\cite{Maas2013} as the activation function. Each FC layer with 128 neurons is followed by a dropout layer~\cite{Srivastava2014}. These layers are then fully connected to the output layer having $n$ neurons with softmax activation function, here $n$ denotes the number of wavelength classes. For attaining maximum classification accuracy over the validation dataset, the above mentioned architecture was chosen after optimizing: the number of DSBs on the range 1 to 10; the number of convolution layers between 1 and 5 for each block; the filter sizes from 1 px $\times$ 1 px to 8 px $\times$ 8 px; and the number of neurons from 8 to 512 by doubling the neurons at each step.

The CNN geometry was optimized by considering the first dataset comprising the speckle images corresponding to 30 different wavelengths at a deviation of 2~fm. This dataset was randomly sampled into 70\% training, 15\% validation and 15\% testing images corresponding to each wavelength. The training was implemented in Matlab 2018a over Nvidia Quadro P5000 GPU. To remove any intensity dependent fluctuations, all the speckle images were zero-center normalized. The CNN was trained to minimize the cross entropy cost function 
\begin{equation}
\text{Cost} = - \frac{1}{k} \Sigma_x [y\times\log a + (1-y)\times\log(1-a)],
\label{Eq1}
\end{equation}
for 10 epochs in the mini batches of 128 images using an ADAM optimizer~\cite{KingmaB14}, where $\Sigma_x$ represents training over all the input images $x$, $k$ is the total number of training data points, $y$ is the target output and $a$ is the network output. Here $y$ and $a$ are the one hot vectors representing the category of the input image. Initial learning rate was set at $1 \times 10 ^{-6}$ and L2 regularization at $2 \times 10^{-4}$. The training process was validated after every 100 iterations.

During the training, the CNN learns to generalize the wavelength-dependent variations of the speckle patterns and thus classify them. The complete CNN architecture can be considered as composed of two ANNs, namely a convolution network (input layer to $FC_1$) and a classifier network ($FC_2$ and $FC_3$). The primary function of the convolution network is to down-sample the 2D input image into a 1D descriptor vector by filtering out the irrelevant / noisy features, whereas the classifier network is trained to classify these 1D descriptor vectors. Thus for a given time instant, where the speckle field is constant with respect to the environmental fluctuations, the convolution network learns to produce a 1D descriptor vector (128 px) corresponding to a particular wavelength for the input 2D image (128 $\times$ 128 px). Hence, after training, the vector output of the convolution network can be directly considered for further analysis. 

\subsection{t-SNE analysis}
\label{tSNE}
To visualize the convolution network's segmentation capabilities over the different datasets, we implemented t-SNE over the generated 1D vectors. t-SNE is a well known non-linear method of machine learning which works on the principle of embedding a low dimensional space such that the neighborhood probabilistic distribution of the higher dimensional data is preserved in the low dimensional vector space. This is achieved by minimizing the symmetric form of Kullback - Leibler divergence~\cite{Maaten2008}. In this study, t-SNE analysis with a perplexity of 30, was implemented using MATLAB 2018b.

\section{Results}
\label{results}
In this section, we present the capabilities of deep learning to measure wavelength deviation from the speckle pattern. Following the training and optimization of the CNN geometry, we identified the limit of detection by showing attometre-scale wavelength precision, a broadband operation range of the CNN based speckle wavemeter, and explored the generalization capabilities of the CNN by changing the disordered medium.

\subsection{CNN optimization}

For the optimization and calibration of the CNN, we varied the wavelength in 2~fm steps over a range of 60~fm. The complete dataset consisted of 10,000 images corresponding to each wavelength, which were randomly divided into training, validation and test datasets by the fraction of 70\%, 15\% and 15\% respectively. To calibrate the CNN, training is performed using every image in the training set.

During the training, a global error is computed by parsing a batch of images sampled randomly from each class of the training set. During the backward pass, this calculated error is backpropagated such that the network can identify the features representing each individual class. In this study, this process is implemented to optimize a CNN architecture over the femtometre-resolved speckle patterns for gaining maximum classification accuracy over the validation dataset.

\begin{figure}[h]
  \centering
  \includegraphics[width=0.8\linewidth]{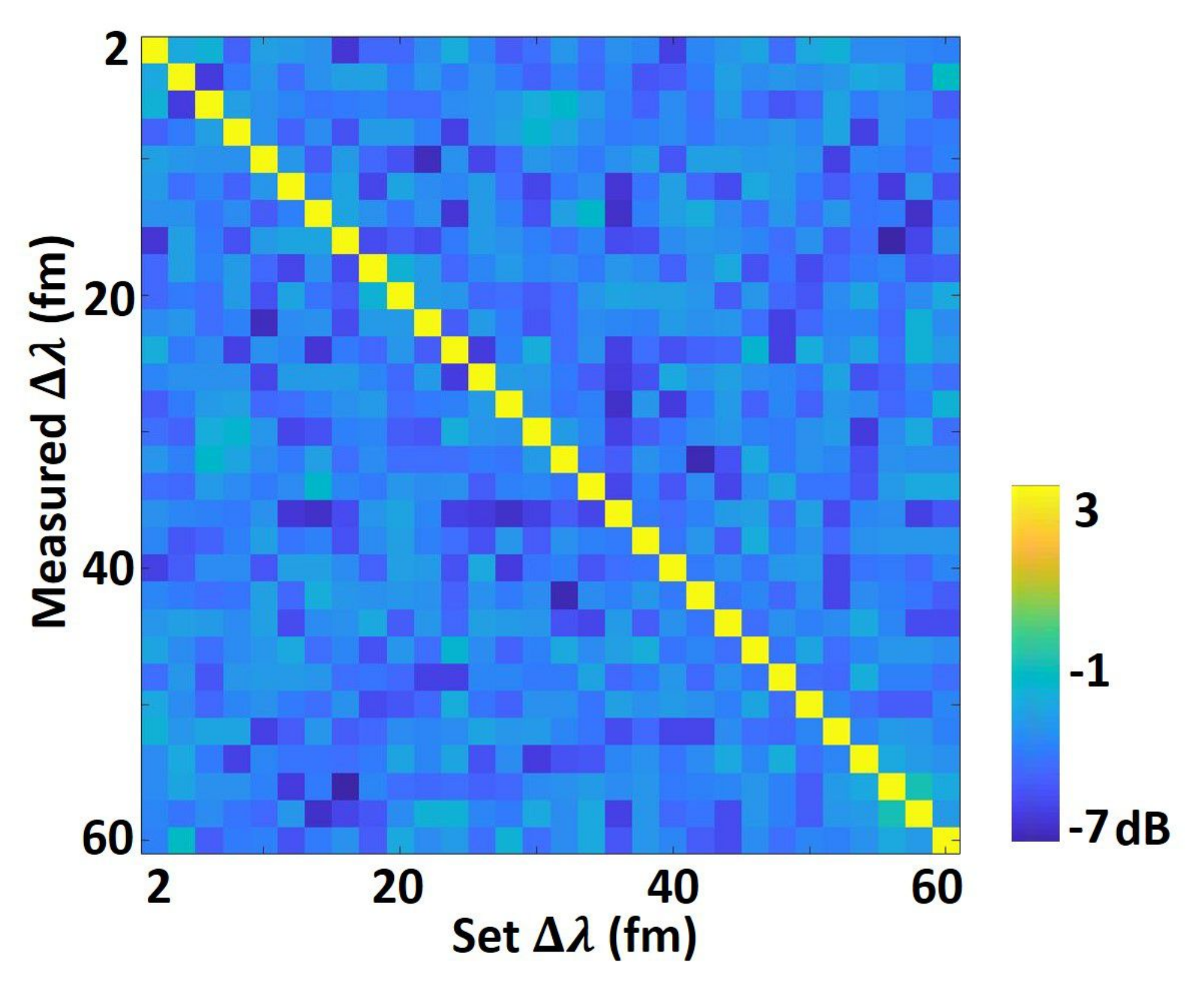}
  \caption{Demonstration of high-accuracy discrimination of femtometre-resolved wavelength changes. Confusion matrix for the output of the CNN in classifying speckle patterns corresponding to wavelength separations on the femtometre-scale, plotted on $\log_{10}$ scale. The color bar represents the decibel values.}
  \label{fig:Fig-2}
\end{figure}

\begin{figure*}[!ht]
  \centering
  \includegraphics[width=1\textwidth]{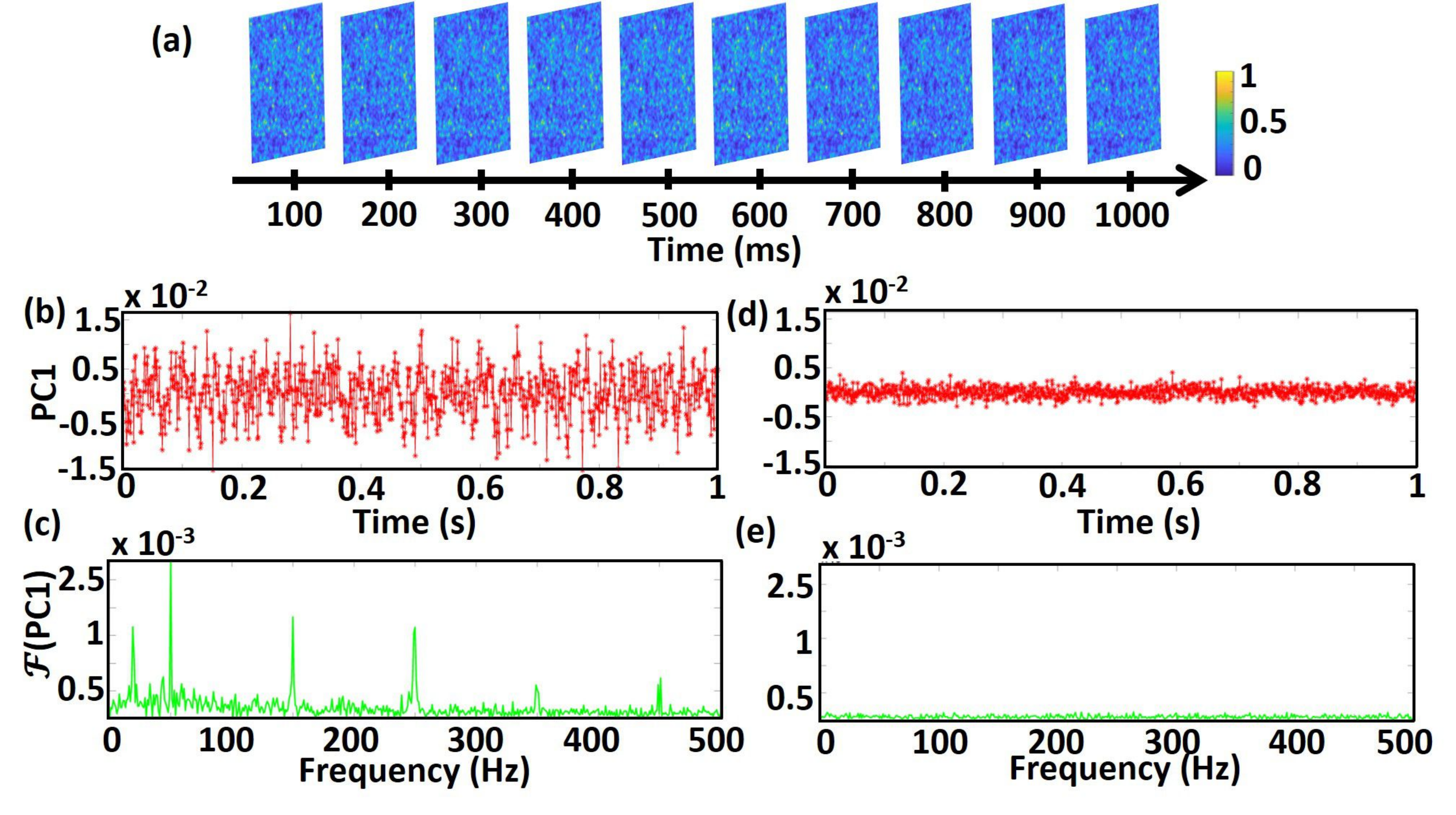}
  \caption{CNN-enabled noise rejection. The CNN learns, through training, to reject instrumental noise from the wavelength measurement. For an example dataset comprising 1000 images sampled over 1 second at a fixed wavelength, (a) shows example speckle images captured at every 100 ms. The color bar represents normalised intensity. (b) PC1 of the full train of speckle images and (c) the Fourier transform of (b) identify the presence of continuous variations in the dataset. (d) PC1 of the 1D descriptor vectors (which are the processed output of the CNN) and (e) the Fourier transform of (d) highlight the absence of any variations present in the output of the CNN. Here, PC1 denotes the first principal component}
  \label{fig:Fig-3}
\end{figure*}

After the training process, images from the test set (a total of 45,000 images across all the wavelengths) were considered for testing the performance of the CNN. The one-hot classification by the CNN led to a very accurate measurement with 100\% classification accuracy. To evaluate the probabilistic classification error of the CNN we calculated the softmax output of the $FC_3$ layer for the test dataset. The error was calculated by taking the sum of all the incorrect classification values for each image and then taking the average of this summed value over all the images. This leads to a probabilistic classification error of $2.2 \times 10^{-6}$. To emphasize this fact, we calculated the confusion matrix from the softmax output of the CNN over $\log_{10}$ scale, which is presented in Fig. \ref{fig:Fig-2}. For the optimal architecture, the CNN classification accuracy does not depend on the number of training classes, or the step size between them.

\subsection{CNN noise rejection capabilities}
It has been shown in a previous study that the CNN, once trained on a given dataset, learns to overcome a low signal to noise environment~\cite{helgadottir2019digital}. In our study we find analogous results. Through the training process, the CNN learns to reject variations in the speckle patterns which do not correspond to the control parameter, i.e., wavelength. This is demonstrated in Fig. \ref{fig:Fig-3}, where we analysed the speckle images by implementing PCA on both the input raw images and the processed output of the CNN. To evaluate an improvement in the stability of the speckle image time series (accumulated for 1 second) before and after the CNN transformation (Fig. \ref{fig:Fig-3} (b) and (d)) we estimated the smallest detectable shift in wavelength to be 3 times the standard deviation ($\sigma$) from the mean position. The 3$\sigma$ value of the first principal component (PC1) was evaluated as 0.014 for the raw speckle images whereas it was evaluated as 0.003 for the output of the CNN, an improvement by a factor of 4.66. Additionally, to analyze the periodic variations in the data, we calculated the Fourier transform of PC1 (Fig \ref{fig:Fig-3} (c) and (e)). For the raw input speckle images at a fixed wavelength, PC1 of the input images show several periodic noise components. However, when the output of the convolution network ($FC_1$) is analysed using PCA, PC1 does not reflect any of the temporal noise components that were present in the input dataset. This shows that the CNN, once trained to classify the speckle images with respect to wavelengths, filters the input speckle images and returns the output as a 1D vector representing a single wavelength without any environmental or instrumental noise.

Given the widely-known capability of ANNs to operate as universal functional approximators~\cite{Hornik1989}, the results in Fig. \ref{fig:Fig-3} (d) and Fig. \ref{fig:Fig-3} (e) also orient us towards a conclusion that once the CNN is trained to classify the speckle images for a single wavelength, it processes the input images to down-sample them into one dimensional vector such that any noisy components are rejected. This suggests that we can further train the CNN to recognise the incident laser wavelength with a precision below the instrumental circuitary noise.

\subsection{Attometre precision}
In order to observe the limit of detection of the trained deep learning model, we accumulated a second dataset where the laser wavelength was tuned over separations on the attometre scale. More specifically, the speckle images were captured by detuning the acousto-optic modulator across five distinct wavelengths with an increment of two attometres. The time taken to accumulate the dataset for a single wavelength was 10 seconds whereas the time difference between the different wavelength classes was typically 0.5 seconds. This ensures that any wavelength drift between measurements should be small compared to the drift within a single measurement period. Moreover, we see no evidence of drift within the measurement period (see Fig \ref{fig:Fig-3}), verifying that each measurement is congruent to a single wavelength.

\begin{figure}[h]
  \centering
  \includegraphics[width=1\linewidth]{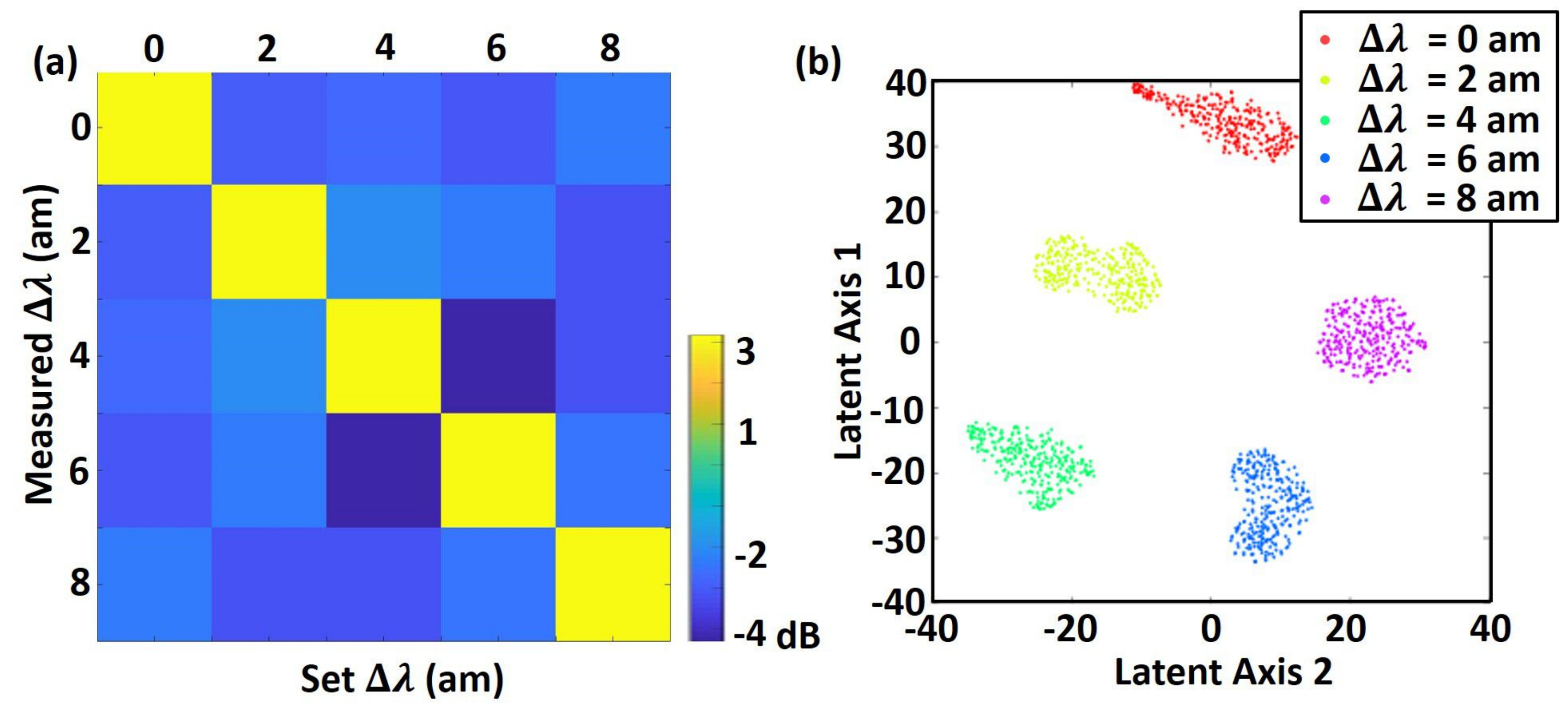}
  \caption{CNN classification and segmentation capabilities of attometre-resolved speckle data. (a) Confusion matrix on the $log_{10}$ scale depicting the classification abilities of the CNN for wavelengths separated by 2~am. Here $\Delta\lambda$ = 0~am depicts a detuning of $\Delta\lambda$ = 30.652~fm from the rubidium crossover and the other values are relative to it. The color bar represents the decibel values. (b) t-SNE visualization for the output of the $FC_1$ from the CNN trained over femtometre-resolved speckle data, applied to speckle data separated by two attometres. The speckle images at each wavelength form a distinct cluster, showing that the CNN can be retrained simply by using a single speckle image at a known wavelength. }
  \label{fig:Fig-5}
\end{figure}

As the dataset is changed, the classification abilities of the CNN needs to be re-tuned, hence by the virtue of transfer learning, we retrained the CNN by changing the number of neurons in the output layer ($FC_3$). A total of 7000 images per wavelength class were considered for training/validation and 3000 images per class were considered for the testing process. The retrained CNN gave a one-hot classification accuracy of 100\% and a softmax probabilistic classification error of $3.8 \times 10^{-5}$. To emphasize the accuracy of measurement, we show the $\log_{10}$ of the confusion matrix in Fig. \ref{fig:Fig-5} (a). This matrix was calculated from the activations of the softmax layer at the output of the trained CNN. These results show that the re-training of a CNN can result in a wavelength precision as low as 2~am. This wavelength precision of two attometres is not a fundamental limit, but limited by the precision with which we can control the wavelength using the AOM in the experiment.

To eliminate the process of retraining the CNN, the images were also processed using the CNN trained to classify the~fm-resolved data. The output 1D vectors at $FC_1$ were further analysed using t-SNE (see section \ref{methods} subsection \ref{tSNE}) to visualize the segmentation capabilities. Figure \ref{fig:Fig-5} (b) denotes the output of $FC_1$ layer of the CNN downsampled to a 2 dimensional latent vector space. Each cluster represents the speckle images corresponding to a specific wavelength. Evidently, using this method we do not need to train the CNN further using 7000 images per class but we can use only a single image for the further classification, and still achieve attometre-scale precision. 

The resolving power of a wavemeter is $R = \frac{\lambda_{0}}{\delta \lambda}$, where $\lambda_{0}$ is the absolute wavelength and $\delta \lambda$ is the minimum detected deviation from it. The resolving power of the deep learning enabled speckle wavemeter is $R > 10^{11}$ for a central wavelength at 780~nm with a least deviation of 2 attometre.

\subsection{Broadband operation range}
We tested the broadband segmentation capabilities of the CNN by accumulating the speckle patterns over two wavelength ranges: from 770~nm to 790~nm in 5~nm increments and separately at 488~nm, 532~nm, 785~nm and 976~nm (see Methods for details).

\begin{figure}[h]
  \centering
  \includegraphics[width=1\linewidth]{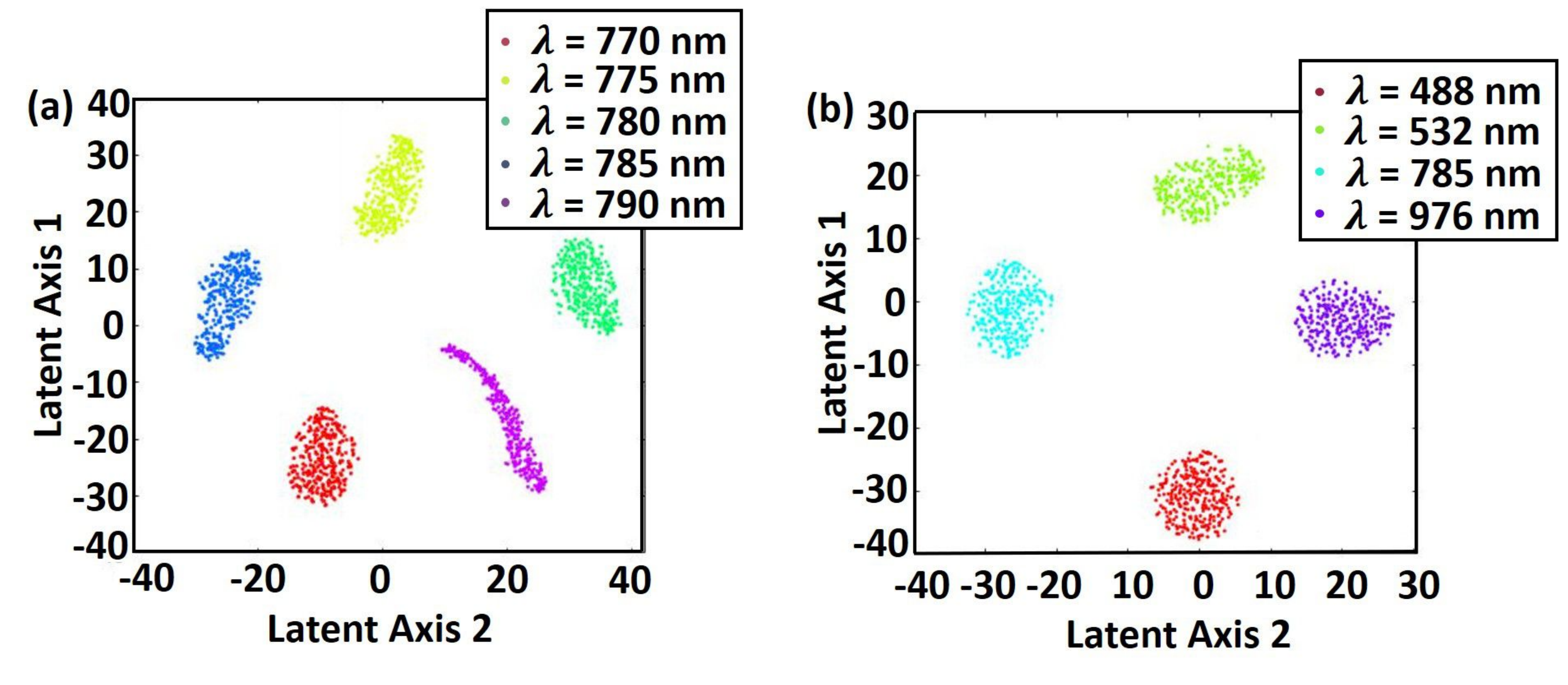}
  \caption{Segmentation capabilities of the CNN over a broadband range of data. (a) t-SNE scatter plot of the output of the 1D descriptor vector for the wavelength deviations corresponding to 770~nm, 775~nm, 780~nm, 785~nm and 790~nm. (b) t-SNE scatter plot of the 1D descriptor vector corresponding to 488~nm, 532~nm, 785~nm and 976~nm.}
  \label{fig:Fig-4}
\end{figure}

When the fm-trained CNN was implemented over the two datasets, and the output of $FC_1$ was analysed using t-SNE, the speckle images corresponding to individual wavelength were clustered independently as depicted in Fig. \ref{fig:Fig-4} (a) and Fig. \ref{fig:Fig-4} (b). With respect to the interpolative estimation, this result shows that once the CNN is trained, it can be harnessed for the classification of speckle images at a broadband range between 488~nm to 976~nm regardless of the variation in the incident laser wavelength.

As before, without retraining the CNN, t-SNE evaluation of the output of $FC_1$ shows a clear clustering of the classes, meaning that a full retraining is not necessary, but wavelength detection can be accomplished simply by using one known wavelength per cluster. The fractional bandwidth of the wavemeter is $B=\left.(\lambda_{max}-\lambda_{min}) \middle/ \frac{1}{2} (\lambda_{max}+\lambda_{min})\right.$, where $\lambda_{max}$ is the maximum detected wavelength and $\lambda_{min}$ is the minimum detected wavelength in the broadband operation range, giving $B = 0.66$ for the speckle wavemeter presented here. 

\subsection{High dynamic range}
Defining the dynamic range as the product $B\times R$, these results showcase the high dynamic range capability of the CNN in classifying the speckle patterns: identifying wavelength differences with a precision of a few attometres over a range of 100s of nanometres gives a high dynamic range of $3.25 \times 10^{11}$.

\begin{figure}[h]
  \centering
  \includegraphics*[width=1\columnwidth]{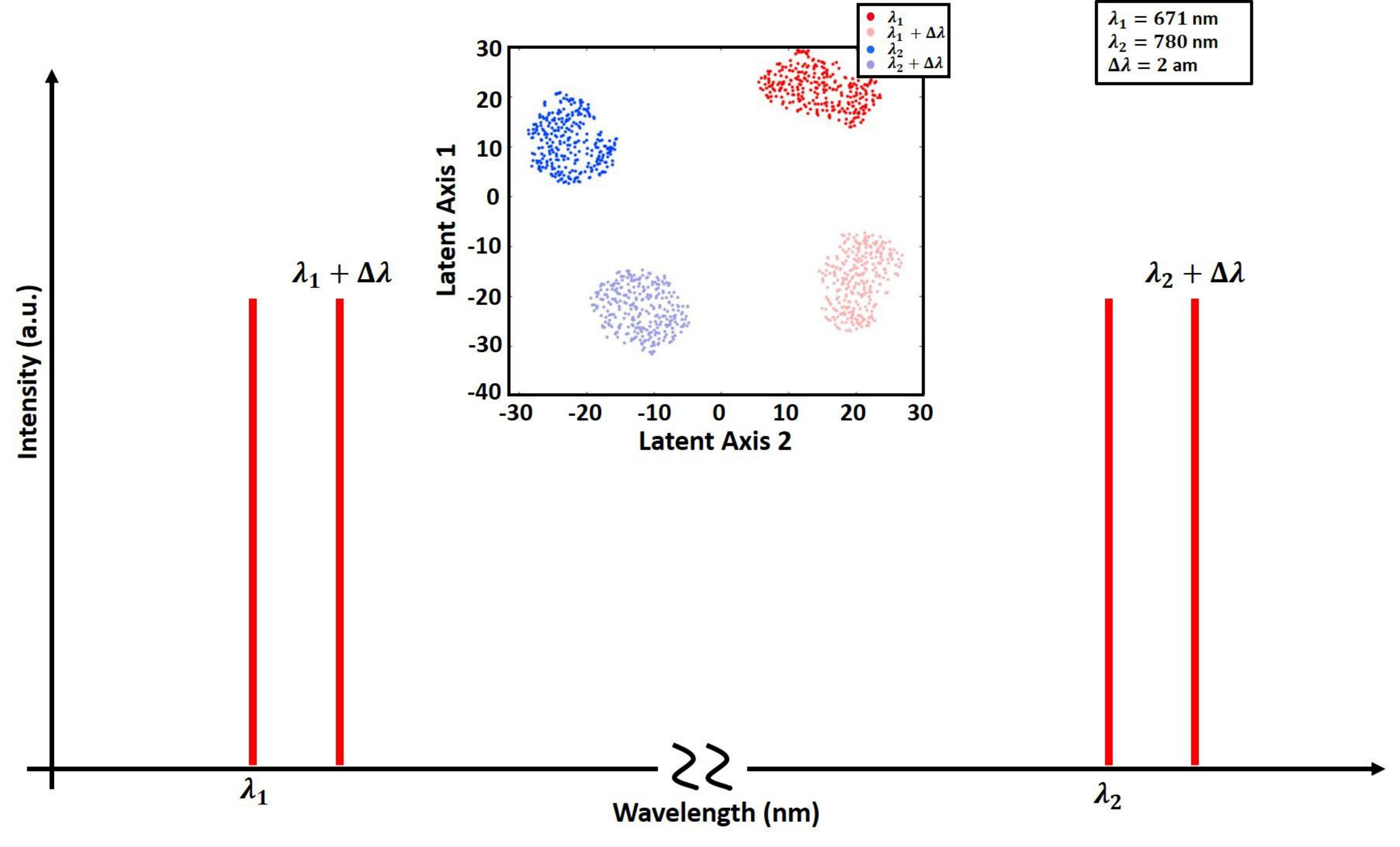}
  \caption{High dynamic range of the CNN. The graph visualizes the speckle patterns accumulated at four distinct wavelengths at 780~nm, 780~nm + $\Delta\lambda$, 671~nm and 671~nm + $\Delta\lambda$. The inset graph represents the t-SNE scatter plot of the output of the 1D descriptor vector for the mentioned wavelengths. (Here $\Delta\lambda = 2$~am)}
  \label{fig:Fig-HDR}
\end{figure}

To display the high dynamic range capabilities of the CNN we also accumulated the data using diode lasers locked to the D2 lines of $^{87}$Rb ($\sim$ 780~nm) and $^{7}$Li ($\sim$ 671~nm). For each laser, we use the AOM to generate two set of speckle patterns, with a wavelength separation of 2~am between each of the two sets. This resulted in a broadband wavelength measurement with a precision of 2~am. As can be inferred from Fig. \ref{fig:Fig-HDR}, t-SNE plot shows the presence of four distinct clusters corresponding to the four mentioned wavelengths.

\subsection{Generalization capabilities of the CNN}

The speckle pattern represents the spatial correlation function of a given disordered medium and the incident light wavelength. This means that if a different scattering medium is used, the speckle pattern would also be different but maintain the wavelength dependent deviations~\cite{George1974}. Therefore, with respect to the generalization capability, a deep learning based model trained to decorrelate the speckle patterns with respect to wavelength should in principle be able to segment the speckle images generated from any random disordered medium. To consider this theory of generalized segmentation properties, we accumulated a dataset using a ground glass diffuser (Thorlabs ED1-S20) in place of the integrating sphere. As can be observed from Fig. \ref{fig:Fig-7} (a) and Fig. \ref{fig:Fig-7} (b), the speckle patterns generated from the integrating sphere and the ground glass show completely different characteristic features. 

We accumulated two sub-datasets, one where the speckle images were accumulated by varying the incident wavelength with an increment of 2~fm and the other with an increment of 20~am. The segmentation capability of the CNN, trained over femtometre resolved data, was tested by processing the speckle images generated from each class. The output from $FC_1$ was analysed using t-SNE and the results are presented in Fig. \ref{fig:Fig-7}. As shown in Fig. \ref{fig:Fig-7} (c) and (d), the CNN segments and clusters each of the speckle images into their individual class.

The results clearly show that the CNN processed the images, which represented completely different spatial variations, and clustered them with respect to the incident laser wavelength.

\section{Discussion}
\label{discussion}

The combination of speckle with CNNs achieves a remarkable classification accuracy since speckle patterns represent an ideal candidate for the training of the CNN. As shown in Fig. \ref{fig:Fig-2}, the CNN achieves a one-hot classification accuracy of 100 \%, with a softmax probabilistic classification error of $2.2 \times 10^{-6}$. If the disordered medium and the laser wavelength are kept constant then, ideally, the resulting speckle pattern should not change. However, the environmental fluctuations or fluctuations due to instrumental circuitry cause the speckle patterns to change with time. Therefore, we have also demonstrated that the CNN, once trained to classify the speckle patterns, automatically learns to reject the environmental or instrumental fluctuations.

\begin{figure}[h]
  \centering
  \includegraphics[width=1\linewidth]{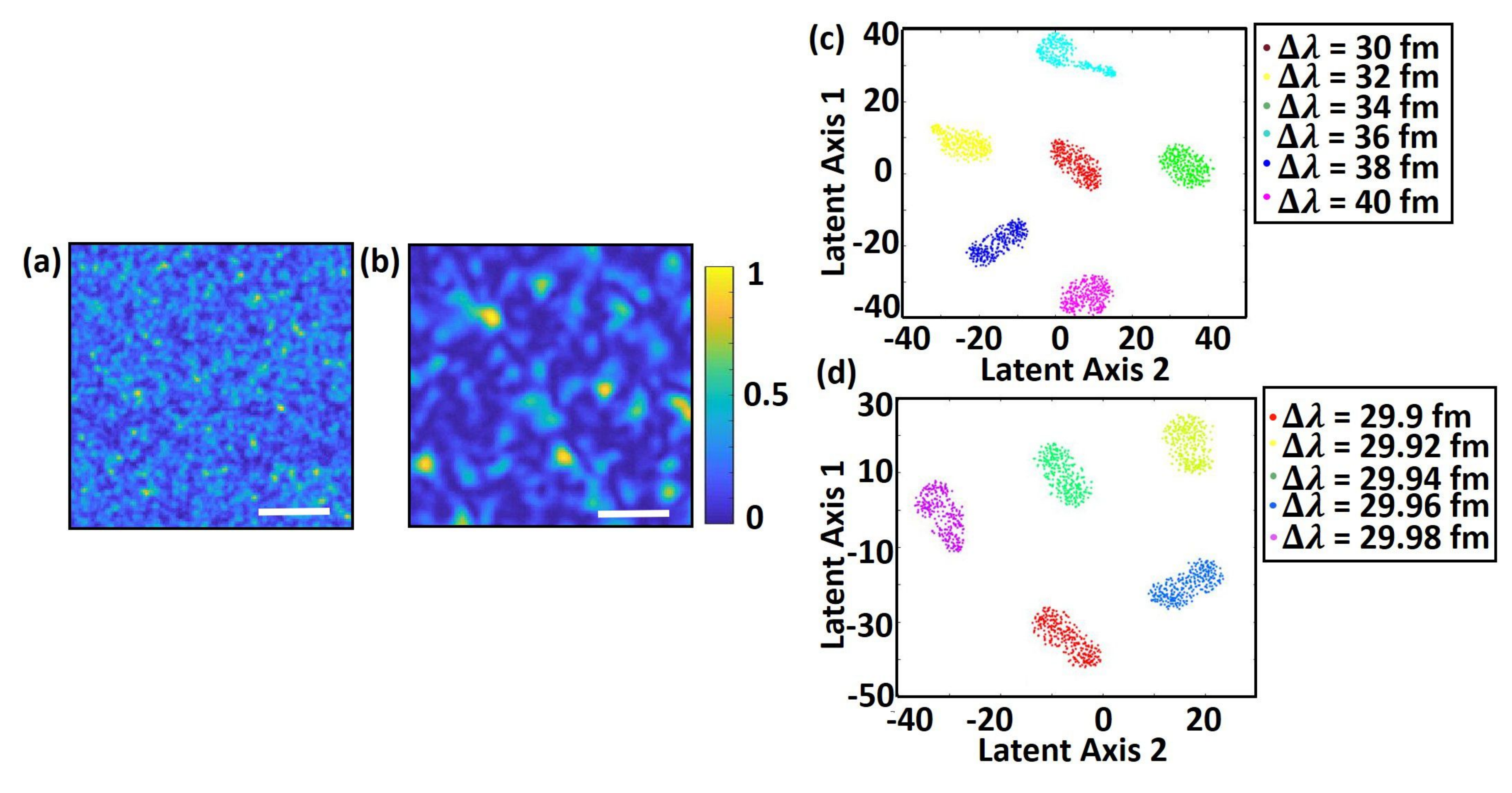}
  \caption{Transferring wavelength classification to a different scattering medium. Speckle pattern generated using an (a) integrating sphere and (b) ground glass assembly. The color bar represents normalised intensity while the white scale bar represents 224~$\mu$m. Segmentation results using ground glass assembly for (c) femtometre-resolved and (d) attometre-resolved incident laser wavelength modulations.}
  \label{fig:Fig-7}
\end{figure}

As can be inferred from Fig. \ref{fig:Fig-3}, the structure of a CNN model and backpropagation training, drives it to progressively learn the filtering of the input images. As explained before, the training is implemented such that the output only contains the features relevant to the individual class of the images in the training dataset. In the case of a speckle wavemeter, conceptually, the speckle images should be down-sampled such that the output only contains the features with respect to the wavelength. In this study, we have demonstrated that the CNNs can accurately classify speckle patterns measured with a wavelength separation of 2~am (Fig. \ref{fig:Fig-5}) which can be attributed to the automated noise rejection capability of the CNN.

The results demonstrated in Fig. \ref{fig:Fig-3}, in combination with the universal function approximator property of the CNN, provide an insight that this model once trained can be implemented (in combination with a dimensionality reduction algorithm) to segment the wavelength-dependent speckle with any deviation and generated from any disordered media. The results presented in Fig. \ref{fig:Fig-4}, Fig. \ref{fig:Fig-HDR} and Fig. \ref{fig:Fig-7} highlight these properties of the deep learning based model.

ANNs, although powerful tools, also come with potential limitations. When compared to other statistical methods, an ANN takes considerably more time for the training process. Moreover, for a classification problem, the training time increases exponentially with the number of classes. Additionally, the ANN which is trained to classify, can only identify data-points which are part of training set classes. This would apparently restrict the range and precision of the speckle wavemeter. Whilst a regression based approach seems to be an attractive option to generalize over the unknown data-points, this generalisation capability seems to be limited at a much lower precision~\cite{Magnusson2018}. Instead, this potential limitation has been overcome here by the application of t-SNE which enables relative wavelength measurement for speckle patterns which are not part of the training set classes.

\section{Conclusion}
\label{conclusion}
In conclusion, this study has implemented a deep learning based method to classify the speckle patterns with respect to the incident light source wavelength. The combination of laser speckle and deep learning provides an accurate method to distinguish between laser wavelengths separated by as little as 2~am and as much as 488~nm, showing a dynamic range of $3.25 \times 10^{11}$ in a single-step algorithm. This combination can also be applied to a completely distinct scattering medium, and re-calibrating the CNN using the method of transfer learning provides an efficient training procedure for a highly accurate wavemeter.

Additionally, this study shows that a trained deep learning model can be implemented to reject inherent instrumental or environmental noise.  The results presented here will be beneficial in not only automated laser stabilization but may also be useful for noise reduction in multiple telecommunication applications. In the future, we will investigate extending this work to the development of robust and compact spectrometers with a capability to measure multiple wavelengths, and will also investigate the possibility to train the CNN to simultaneously measure wavelength, polarisation and transverse mode profile.

\section*{Acknowledgement}
The authors would like to acknowledge technical assistance from Dr. Donatella Cassettari. This work was supported by a Medical Research Scotland PhD studentship PhD 873-2015 awarded to R.K.G, and grant funding from Leverhulme Trust (RPG-2017-197) and UK Engineering and Physical Sciences Research Council (grant EP/P030017/1). The opinions expressed in this article are the authors own and do not reflect the view of above mentioned funding agencies. The experiments were designed by RKG, GDB and KD. RKG  developed and designed the CNN and performed the numerical analysis. GDB developed the experimental setup and performed the experiments. KD and SJP supervised the study. The manuscript was written by RKG with contributions from GDB and KD. All authors approved the manuscript. Research data supporting this publication can be accessed at \url{https://doi.org/10.17630/d6049ec8-972e-4820-9cd4-9552b26c8426}

\bibliography{scibib}

\begin{thebibliography}{10}

\bibitem{PhysRevA.92.052501}
R.~H. Leonard, A.~J. Fallon, C.~A. Sackett, and M.~S. Safronova.
\newblock High-precision measurements of the $^{87}\mathrm{Rb}$ ${D}$-line
  tune-out wavelength.
\newblock {\em Phys. Rev. A}, 92:052501, 2015.

\bibitem{erf2012speckle}
Robert Erf.
\newblock {\em Speckle Metrology}.
\newblock Elsevier, 2012.

\bibitem{Chakrabarti2015}
Maumita Chakrabarti, Michael~Linde Jakobsen, and Steen~G. Hanson.
\newblock {Speckle-based spectrometer}.
\newblock {\em Opt. Lett.}, 40:3264, 2015.

\bibitem{cao2017perspective}
Hui Cao.
\newblock Perspective on speckle spectrometers.
\newblock {\em J. Opt.}, 19:060402, 2017.

\bibitem{Redding2012}
Brandon Redding and Hui Cao.
\newblock {Using a multimode fiber as a high-resolution, low-loss
  spectrometer}.
\newblock {\em Opt. Lett.}, 37:3384, 2012.

\bibitem{Redding2013}
Brandon Redding, Sebastien~M. Popoff, and Hui Cao.
\newblock {All-fiber spectrometer based on speckle pattern reconstruction}.
\newblock {\em Opt. Express}, 21:6584, 2013.

\bibitem{Redding2014}
Brandon Redding, Mansoor Alam, Martin Seifert, and Hui Cao.
\newblock {High-resolution and broadband all-fiber spectrometers}.
\newblock {\em Optica}, 1:175, 2014.

\bibitem{Wan2015}
Noel~H. Wan, Fan Meng, Tim Schr{\"{o}}der, Ren-Jye Shiue, Edward~H. Chen, and
  Dirk Englund.
\newblock {High-resolution optical spectroscopy using multimode interference in
  a compact tapered fibre}.
\newblock {\em Nat. Commun.}, 6:7762, 2015.

\bibitem{Bruce:19}
Graham~D. Bruce, Laura O'Donnell, Mingzhou Chen, and Kishan Dholakia.
\newblock {Overcoming the speckle correlation limit to achieve a fiber
  wavemeter with attometer resolution}.
\newblock {\em Opt. Lett.}, 44(6):1367--1370, 2019.

\bibitem{Bruce2020}
Graham~D Bruce, Laura O’Donnell, Mingzhou Chen, Morgan Facchin, and Kishan
  Dholakia.
\newblock Femtometer-resolved simultaneous measurement of multiple laser
  wavelengths in a speckle wavemeter.
\newblock {\em Opt. Lett.}, 45(7):1926--1929, 2020.

\bibitem{Metzger2017}
Nikolaus~Klaus Metzger, Roman Spesyvtsev, Graham~D. Bruce, Bill Miller,
  Gareth~T. Maker, Graeme Malcolm, Michael Mazilu, and Kishan Dholakia.
\newblock {Harnessing speckle for a sub-femtometre resolved broadband wavemeter
  and laser stabilization}.
\newblock {\em Nat. Commun.}, 8:15610, 2017.

\bibitem{ODONNELL2020124906}
Laura O'Donnell, Kishan Dholakia, and Graham~D. Bruce.
\newblock High speed determination of laser wavelength using poincaré
  descriptors of speckle.
\newblock {\em Opt. Commun.}, 459:124906, 2020.

\bibitem{PhysRevLett.104.100601}
S.~M. Popoff, G.~Lerosey, R.~Carminati, M.~Fink, A.~C. Boccara, and S.~Gigan.
\newblock Measuring the transmission matrix in optics: An approach to the study
  and control of light propagation in disordered media.
\newblock {\em Phys. Rev. Lett.}, 104:100601, 2010.

\bibitem{Mazilu2014}
Michael Mazilu, Tom Vettenburg, Andrea {Di Falco}, and Kishan Dholakia.
\newblock {Random super-prism wavelength meter}.
\newblock {\em Opt. Lett.}, 39:96, 2014.

\bibitem{6472238}
Y.~{Bengio}, A.~{Courville}, and P.~{Vincent}.
\newblock Representation learning: A review and new perspectives.
\newblock {\em IEEE Transactions on Pattern Analysis and Machine Intelligence},
  35(8):1798--1828, Aug 2013.

\bibitem{10.1126/sciadv.aaw4967}
Jennifer~F. Hoyal~Cuthill, Nicholas Guttenberg, Sophie Ledger, Robyn Crowther,
  and Blanca Huertas.
\newblock Deep learning on butterfly phenotypes tests
  evolution{\textquoteright}s oldest mathematical model.
\newblock {\em Sci. Adv.}, 5(8):eaaw4967, 2019.

\bibitem{PhysRevLett.122.040504}
Maria Schuld and Nathan Killoran.
\newblock Quantum machine learning in feature hilbert spaces.
\newblock {\em Phys. Rev. Lett.}, 122:040504, 2019.

\bibitem{Gupta:19}
Roopam~K. Gupta, Mingzhou Chen, Graeme P.~A. Malcolm, Nils Hempler, Kishan
  Dholakia, and Simon~J. Powis.
\newblock Label-free optical hemogram of granulocytes enhanced by artificial
  neural networks.
\newblock {\em Opt. Express}, 27(10):13706--13720, May 2019.

\bibitem{Hornik1989}
Kurt Hornik, Maxwell Stinchcombe, and Halbert White.
\newblock {Multilayer feedforward networks are universal approximators}.
\newblock {\em Neural Networks}, 2:359--366, 1989.

\bibitem{Baldi2014}
P.~Baldi, P.~Sadowski, and D.~Whiteson.
\newblock Searching for exotic particles in high-energy physics with deep
  learning.
\newblock {\em Nat. Commun.}, 5:4308, 2014.

\bibitem{Rem2019}
Benno~S. Rem, Niklas K{\"a}ming, Matthias Tarnowski, Luca Asteria, Nick
  Fl{\"a}schner, Christoph Becker, Klaus Sengstock, and Christof Weitenberg.
\newblock Identifying quantum phase transitions using artificial neural
  networks on experimental data.
\newblock {\em Nat. Phys.}, 15:917--920, 2019.

\bibitem{wang2019deep}
Hongda Wang, Yair Rivenson, Yiyin Jin, Zhensong Wei, Ronald Gao, Harun
  G{\"u}nayd{\i}n, Laurent~A Bentolila, Comert Kural, and Aydogan Ozcan.
\newblock Deep learning enables cross-modality super-resolution in fluorescence
  microscopy.
\newblock {\em Nat. Methods}, 16:103--110, 2019.

\bibitem{Horisaki2018}
Ryoichi Horisaki, Ryosuke Takagi, and Jun Tanida.
\newblock {Deep-learning-generated holography}.
\newblock {\em Appl. Opt.}, 57:3859, 2018.

\bibitem{Li2018}
Yunzhe Li, Yujia Xue, and Lei Tian.
\newblock {Deep speckle correlation: a deep learning approach toward scalable
  imaging through scattering media}.
\newblock {\em Optica}, 5:1181, 2018.

\bibitem{Valent2018}
Eadan Valent and Yaron Silberberg.
\newblock {Scatterer recognition via analysis of speckle patterns}.
\newblock {\em Optica}, 5:204, 2018.

\bibitem{Kurum:19}
Ulas K\"{u}r\"{u}m, Peter~R. Wiecha, Rebecca French, and Otto~L. Muskens.
\newblock Deep learning enabled real time speckle recognition and hyperspectral
  imaging using a multimode fiber array.
\newblock {\em Opt. Express}, 27:20965--20979, 2019.

\bibitem{Maaten2008}
Laurens van~der Maaten and Geoffrey Hinton.
\newblock {Visualizing Data using t-SNE}.
\newblock {\em J. Mach. Learn. Res.}, 9:2579--2605, 2008.

\bibitem{Facchin20}
M.~Facchin, G.~D. Bruce, and K.~Dholakia.
\newblock Speckle-based determination of the polarisation state of single and
  multiple laser beams.
\newblock {\em OSA Continuum}, 3:1302, 2020.

\bibitem{Mourka13}
A~Mourka, M~Mazilu, Ewan~M Wright, and K~Dholakia.
\newblock Modal characterization using principal component analysis:
  application to bessel, higher-order gaussian beams and their superposition.
\newblock {\em Sci. Rep.}, 3:1422, 2013.

\bibitem{reddy2014higher}
Salla~Gangi Reddy, Shashi Prabhakar, Ashok Kumar, J~Banerji, and RP~Singh.
\newblock Higher order optical vortices and formation of speckles.
\newblock {\em Opt. Lett.}, 39(15):4364--4367, 2014.

\bibitem{hu2020does}
Xiao-Bo Hu, Meng-Xuan Dong, Zhi-Han Zhu, Wei Gao, and Carmelo
  Rosales-Guzm{\'a}n.
\newblock Does the structure of light influence the speckle size?
\newblock {\em Sci. Rep.}, 10:199, 2020.

\bibitem{Maas2013}
Andrew~L. Maas, Awni~Y. Hannun, and Andrew~Y. Ng.
\newblock {Rectifier Nonlinearities Improve Neural Network Acoustic Models}.
\newblock In {\em Proc. ICML}, volume~30, page~3, 2013.

\bibitem{Srivastava2014}
Nitish Srivastava, Geoffrey Hinton, Alex Krizhevsky, Ilya Sutskever, and Ruslan
  Salakhutdinov.
\newblock {Dropout: A Simple Way to Prevent Neural Networks from Overfitting}.
\newblock {\em J. Mach. Learn. Res.}, 15:1929--1958, 2014.

\bibitem{KingmaB14}
Diederik~P. Kingma and Jimmy Ba.
\newblock Adam: {A} method for stochastic optimization.
\newblock In {\em 3rd International Conference on Learning Representations,
  {ICLR} 2015, San Diego, CA, USA, May 7-9, 2015, Conference Track
  Proceedings}, 2015.

\bibitem{helgadottir2019digital}
Saga Helgadottir, Aykut Argun, and Giovanni Volpe.
\newblock Digital video microscopy enhanced by deep learning.
\newblock {\em Optica}, 6(4):506--513, Apr 2019.

\bibitem{George1974}
Nicholas George and Atul Jain.
\newblock {Space and wavelength dependence of speckle intensity}.
\newblock {\em Appl. Phys.}, 4:201--212, 1974.

\bibitem{Magnusson2018}
Einar~B. Magnusson, J.~P.~Balthasar Mueller, Michael Juhl, Carlos Mendoza, and
  Kristjan Leosson.
\newblock {Neural Polarimeter and Wavemeter}.
\newblock {\em ACS Photonics}, 5:2682--2687, 2018.

\end{thebibliography}

\bibliographystyle{unsrt}

\end{document}